%% file: p-Conf-99-204-E.tex
\documentstyle[preprint,aps,epsf]{revtex} 

\def\NIM{Nucl. Instrum. Methods}
\def\NP{Nucl. Phys.}
\def\PRL{Phys. Rev. Lett.}
\def\PRD{Phys. Rev. D}
 
\def\ppbar{$p\overline{p}~$}             
\def\pbarp{$\overline{p}p~$}             
\def\qqbar{$q\overline{q}~$}             
\def\pt{$p_T~$}                          
\def\met{\mbox{${\hbox{$E$\kern-0.6em\lower-.1ex\hbox{/}}}_T~$}} 
\def\D0{D\O}                            
\def\etal{{\sl et al.}}                 
\newcommand{\Z}{{\it Z}}

\newcommand{\Dzero}{D\O\ }
\newcommand{\ie}{{\rm i.e.}}

\newcommand{\Wev}{$W\rightarrow e\nu$}

\def\Zee{\mbox{$Z \rightarrow e e \ $}}
\input psfig.tex
\begin{document}
\lefthyphenmin=2
\righthyphenmin=3

%
%
\title{Measurement of the Transverse Momentum Distributions of $W$ and $Z$ Bosons
Produced in \pbarp Collisions at $\sqrt{s} = 1.8\;\rm TeV$}

\author{\centerline{The D\O\ Collaboration
  \thanks{Submitted to the {\it International Europhysics Conference} on
         {\it High Energy Physics}, {\it EPS-HEP99},
          \hfill\break
           15 -- 21 July, 1999, Tampere, Finland.}}}
\address{
\centerline{Fermi National Accelerator Laboratory, Batavia, Illinois 60510}
}

%
%
\date{\today}

\maketitle

%
%
\begin{abstract}
We present measurements of the transverse momentum distribution of $W$ and $Z$ bosons 
produced in {\mbox{$p\bar p$}}\ collisions at {\mbox{$\sqrt{s}$ =\ 1.8\ TeV}}.
The data were collected with the D\O\ detector at Fermilab during 1994-1996 
Tevatron run.
The results are in good agreement with theoretical predictions 
based on the perturbative QCD and soft gluon resummation combined calculation 
over the entire measured \pt range ($p_T = 0-200$~GeV$/c$).

\end{abstract}

\newpage
\begin{center}
\input{list_of_authors}
\end{center}

\normalsize

\vfill\eject

\section{Introduction}
\label{sec:intro}
At the Fermilab Tevatron, 
$W$ and $Z$ bosons
are produced in high energy \pbarp collisions.
The study of the production of $W$ and $Z$ bosons
provides an avenue to explore QCD, the theory of strong
interactions.  
The benefits of using intermediate vector bosons to
study perturbative QCD are large momentum transfer, distinctive
event signatures, low backgrounds, and a well understood electroweak
vertex.  
In this paper we present measurements of the $W$ and $Z$ transverse momentum
distribution based on data taken by the \D0 collider
detector during the 1994--1996 Tevatron running period.  

In the parton model at lowest order, $W$ and $Z$ intermediate vector bosons
are produced in head-on collisions of \qqbar constituents of the 
proton and antiproton, and have little transverse momentum
($p_T << M_W$, $M_Z$).
Consequently, the fact that observed bosons have large 
transverse momentum ($p_T$) is attributed to the production of
one or more gluons or quarks along with the bosons.
At high transverse momentum ($p_T > 20\;\rm GeV/\it c$), 
the cross section is dominated by the
radiation of a single parton with large transverse momentum.
Perturbative QCD is therefore expected to be reliable in
this regime~\cite{ARtheory}.  
At low transverse momentum ($p_T < 10\;\rm GeV/\it c$),
multiple soft gluon emission is expected to dominate the cross section. 
A soft gluon resummation technique~\cite{AKtheory,CSS,DWS,LY} 
is therefore used to make QCD predictions.  
Neither the resummed nor the fixed order calculation describes the distribution
for all values of $p_T$. 
Conventionally, one switches from the resummed calculation to the fixed-order calculation
at $p_T\approx Q$~\cite{AKtheory,LY}.
Thus, a measurement of the transverse momentum distribution may be used to
check the soft gluon resummation calculations in the low \pt
range, and to test the perturbative QCD calculations at high $p_T$.

\section{Introduction to theory}

In standard fixed order perturbative QCD, the partonic cross
section is calculated by expanding in terms of the strong coupling constant
($\alpha_s$)\cite{ARtheory}, where each power of $\alpha_s$
corresponds to the radiation of a single gluon or quark into the
final state. 
This procedure works well in the high \pt region, 
where $p_{\rm T}^2 \approx Q^2$.
However, as
$p_{\rm T} \rightarrow 0$, this fixed order calculation of the
cross section diverges
due to the presence of correction terms that go as $\log^n(Q^2/p_{\rm T}^2)$.
Physically, this failure of fixed order perturbation theory at 
low \pt is due to soft gluon radiation from the initial partons that
is not properly accounted for in the standard expansion.
This difficulty in performing the calculation
can be remedied by reordering the perturbation series through a technique
called resummation, where
the cross section is calculated in terms of the large logarithms, 
rather than strictly in terms of powers of $\alpha_s$.
The resulting calculation is an all orders calculation, \ie,
each piece in the new sum contains terms to all orders in $\alpha_s$. 
However, the largest terms dropped in the latest calculation
are $O(\alpha_s^2)$, so it is considered to be accurate to $O(\alpha_s^2)$.

In final form, the calculation is carried
out via a Fourier transform in impact parameter space ($b$-space),
with the following relation describing  the differential
cross section~[2-5]:
\begin{equation}
$$\frac{\displaystyle d^2\sigma}{\displaystyle dp_{\rm T}^2dy} \sim \int_0^{\infty}d^2b\,
e^{i\vec p_{\rm T}\cdot \vec b}\,W(b,Q) + Y(b,Q)$$
\label{eqvbpty}
\end{equation}
where $W(b,Q)$ contains the results of resumming the perturbative series,
and $Y(b,Q)$ adds back to the calculation the pieces that are perturbative
in $\alpha_s$, but not singular at $p_{\rm T}=0$~\cite{CSS}.

Although the resummation technique extends the applicability of perturbative QCD to lower
values of $p_{\rm T}$, a more fundamental barrier is encountered when $p_{\rm T}$
approaches $\Lambda_{QCD}$.
In this region non-perturbative aspects of the strong force dominate the
production of vector bosons and in general perturbative QCD is expected to fail.
This implies that the resummed calculation becomes undefined above some value of
$b=b_{max}$.
In order to extend the calculation to the low \pt region
a parameterization which accounts for the non-perturbative
effects is introduced.
This extension is accomplished by cutting off the integral in
Eq. \ref{eqvbpty} at some value $b_{max}$ and replacing $W(b,Q)$ with
$W(b_* ,Q)e^{-S_{NP}(b,Q)}$, where
$b_* = b/\sqrt{1+b/b_{max}}$.
This effectively cuts off the contribution of $W(b,Q)$ near $b_{max}$, 
leaving the cross section dominated by the function being introduced 
$S_{NP}(b,Q)$, known as the {\it non-perturbative Sudakov form factor}. 
$S_{NP}$ has the generic renormalization group invariant form~\cite{CSS}:
\begin{equation}
$$S_{NP}(b,Q)=h_1(b,x_A)+h_1(b,x_B)+h_2(b)\ln\left({Q^2\over Q_o^2}\right)$$
\end{equation}
where $x_A$ and $x_B$ are the momentum fractions of the incoming partons, 
$b$ is Fourier conjugate to the transverse momentum (impact parameter), 
$Q_o$ is an arbitrary momentum scale and
$h_1(b,x)$ and $h_2(b)$ are phenomenological functions to be 
determined from experiment~\cite{AKtheory,DWS,LY}.
We used the Sudakov factor functional form from Ladinsky and Yuan~\cite{LY}:
\begin{equation}
$$S^{LY}_{NP}(b,Q)=g_1b^2+g_2b^2\ln\left({Q^2\over Q_o^2}\right)+g_3b\ln(100x_Ax_B)$$
\label{eqly}
\end{equation}
The values of $g_i$ were determined by Ladinsky and Yuan 
by fitting to low energy Drell-Yan data and a small
sample of $\Zee$ data from 1994-96 run at CDF~\cite{CDF}, yielding
$g_1 = 0.11^{+0.04}_{-0.03}\;\rm GeV^2$, $g_2 = 0.4_{-0.2}^{+0.1}\;\rm GeV^2$ and
$g_3 = -1.5_{-0.2}^{+0.1}\;\rm  GeV^{-1}$,
where $b_{max}=0.5\;\rm GeV^{-1}$ and $Q_o=1.6\;\rm GeV$,
and the CTEQ2M pdfs were used.

\section {Measurement of the differential 
 Cross Sections}

In this paper we present measurements of the \pt spectra of
$W$ and $Z$ bosons produced in {\mbox{$p\bar p$}}\ collisions at
{\mbox{$\sqrt{s}$ =\ 1.8\ TeV}} with the D\O\ detector~\cite{D0detector}
at Fermilab. 
The transverse momentum spectra of $W$ and $Z$ bosons have been
measured previously by the UA1~\cite{UA1}, UA2~\cite{UA2},
CDF~\cite{CDF} and D\O\ \cite{wptprl} collaborations, but with smaller data samples than
the ones reported on here.

The $W$ and $Z$ samples have been selected from data taken during the 1994-96 run of the
Tevatron, and correspond to an integrated luminosity of 
$\approx 80\;\rm pb^{-1}$ for $W$ and
$\approx 110\;\rm pb^{-1}$ for $Z$. 
The measurements of the $W$ and $Z$ boson \pt spectra used the
decay modes $W\rightarrow e\nu$ and $Z\rightarrow e^+ e^-$.
Electrons were detected in hermetic, uranium liquid-argon calorimeters
with an energy resolution of about $15\%/\sqrt{E (\mbox{GeV})}$.
The calorimeters have a transverse granularity of $\Delta\eta \times
\Delta\phi = 0.1 \times 0.1$, where $\eta$ is the pseudorapidity
and $\phi$ is the azimuthal angle.
Electrons were accepted in the regions $|\eta|<1.1$ (central) and
$1.5<|\eta|<2.6$ (forward).
In reconstructing the \pt of the $W$ boson, we assume that
the transverse momentum of the neutrino is given by
the calorimetric measurement of the missing
transverse energy 
(\met) in the event.
Electrons from $W$ and $Z$ boson decays tend to be isolated.
Thus, we required the cut
$$\frac{E_{tot}(0.4)-E_{EM}(0.2)}{E_{EM}(0.2)}<0.15,$$
where $E_{tot}(0.4)$ is the energy within $\Delta R < 0.4$ of the cluster
centroid ($\Delta R = \sqrt{\Delta \eta^2 + \Delta \phi^2}$) and $E_{EM}(0.2)$
is the energy in the EM calorimeter within $\Delta R < 0.2$.

At trigger level, the $Z$ analysis required two
electrons, one with transverse energy ($E_T$) greater than 20 GeV 
and the second with $E_T$ greater than 16 GeV.
Off-line for a ``loose" electron we required that $E_T > 25$ GeV and
that the transverse and longitudinal shower shapes be
consistent with those expected for an electron (based on test beam
measurements).
For a ``tight" electron we also required a good
match between a reconstructed track in the drift
chamber system and the shower position in the calorimeter.
For the $Z$ boson sample we required one electron to be ``tight'' and the
other to be either ``tight'' or ``loose''; at least one of the two
electrons had to be in the central region.  
To be acceptable candidates for $Z$ production, both decay electrons 
are required to be isolated.
The dielectron invariant mass was required to lie in the range $75-105$ GeV/c$^2$.
These selection cuts were passed by 6407 $Z$ candidates.

In the case of the $W$ boson analysis, a single electron with
$E_T$ greater than 20 GeV was required at trigger level.
Off-line, a tighter requirement on the electron quality was introduced to reduce the
background level from QCD events, especially at high transverse momentum.
Electron identification was based on a likelihood technique.
Candidates were first identified by finding isolated clusters of energy in the EM 
calorimeter with a matching track in the central detector.
We then cut on a likelihood constructed from the following four variables:
the $\chi^2$ from a covariance matrix which measures the consistency of the calorimeter
cluster shape with that of an electron shower;
the electromagnetic energy fraction (defined as the ratio of the portion of the 
energy of the cluster found in the EM calorimeter to its total energy); 
a measure of the consistency between the track position and the cluster centroid;
and the ionization $dE/dx$ along the track.
For the $W$ boson sample we
required one isolated electron in the central region and \met$>25$ GeV.
The event is rejected if there is a second electron and the 
dielectron invariant mass lies in the range $75-105$ GeV/c$^2$.
A total of 41173 $W$ candidates passed these cuts in the central region.
 
The absolute normalization of the trigger and selection efficiencies were
determined using 
{\mbox{$ Z\rightarrow {e^+e^-}$}}\ \D0 collider events in which one of the electrons
satisfied the trigger and selection criteria. The second electron then
provided an unbiased sample with which to measure the efficiencies.
The variation in the electron selection efficiency as a function of
\pt has been determined using a signal Monte Carlo sample from HERWIG,
smeared for detector resolutions and overlayed with events taken in random \ppbar collisions
(zero bias).  
A parametric Monte Carlo program~\cite{wmassprd}
was used to simulate the \D0 detector response and
calculate the kinematic and geometric acceptance
as a function of $p_T$.
The detector resolutions used in the
Monte Carlo were determined from data, and were parameterized as
a function of energy and angle. The relative response of the hadronic and EM
calorimeters was studied using the transverse momentum of the $Z$ boson
as measured by the \pt of the two electrons compared to the hadronic recoil
system in the $Z$ event.
This parameterized representation of the \D0 detector was
used to smear the predictions by detector effects and
compare it to our measured results.

\begin{figure}[t]
\begin{tabular}{l l}
\psfig{figure=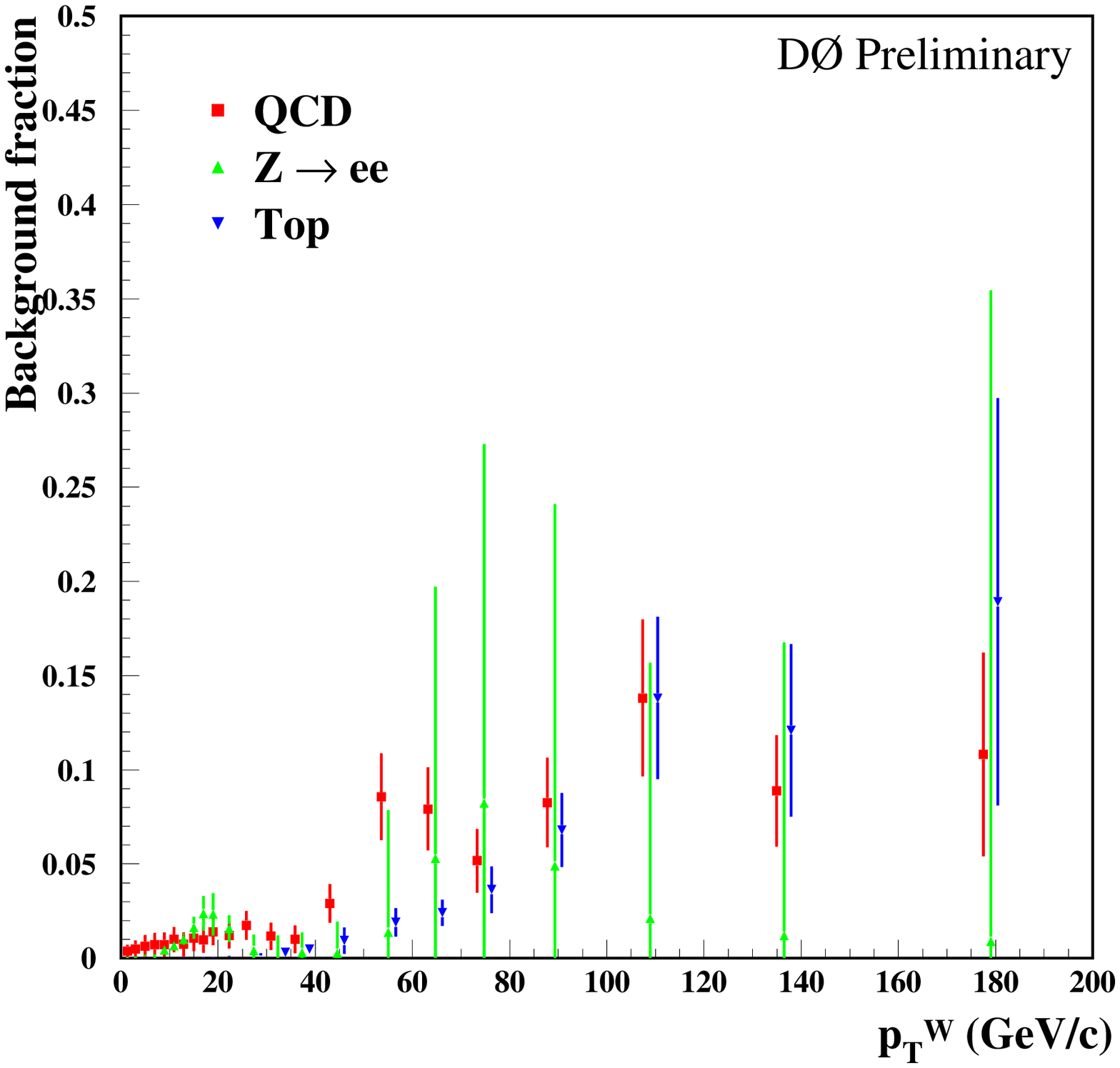,height=8.2cm,width=8.2cm}
&
\psfig{figure=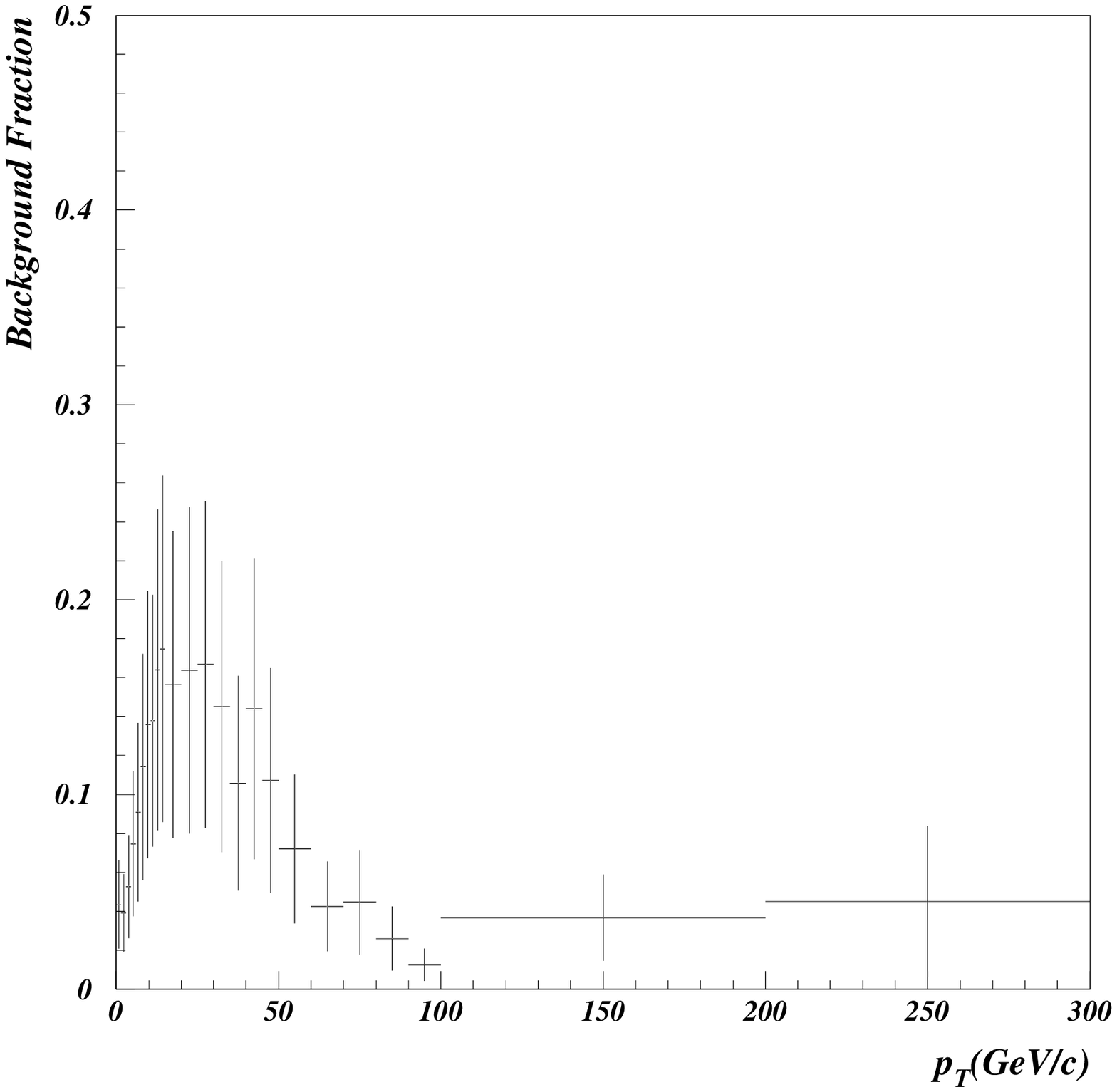,height=7cm,width=7cm}
\end{tabular}
\caption{Left: The background fraction to the \Wev\ signal
from QCD, $\Zee$, and top production as a function of the \pt
of the $W$. Right: The background fraction
from QCD processes to the $\Zee$ signal
as a function of the \pt of the $Z$.}
\label{ptback}
\end{figure}

The major source of background for both the $W$ and the $Z$ sample 
is QCD multi--jet and photon--jet events:
the amount of background in the samples and its shape as a function of \pt was
obtained directly from \D0 data. 

The total amount of QCD background in the inclusive $W$ sample is $\approx 1\%$.
The normalized QCD background
was subtracted bin-by-bin from the $W$ boson candidate sample transverse
momentum spectrum.
Additional corrections were made to account for
top quark background events (0.1\%) and for
{\mbox{$ Z\rightarrow {e^+e^-}$}}\ events (0.8\%), where one of the
electrons was lost or not identified.
Since $p_T^{W}$ was measured from the recoiling hadrons,
the events originating from
$W\rightarrow \tau\nu$ (where $\tau\rightarrow e \nu\nu$)
contributed properly to the differential distribution; this source
of background therefore was included in the Monte Carlo simulation of the
$p_T^{W}$ distribution.

The total background for the $Z$ sample is $\approx 2\%$ for events in
which both electrons are in the central region and $\approx 7\% $ when one electron
is in the central region and the other is in the forward region. 
Backgrounds to $\Zee$ production from
$Z\rightarrow \tau \tau$, top quark and diboson production have been
estimated from Monte Carlo samples and are negligible. 
Figure \ref{ptback} shows the background fractions as a function of \pt
for both the $W$ and $Z$ samples.

\begin{figure}[ht]
\begin{tabular}{l l}
\psfig{figure=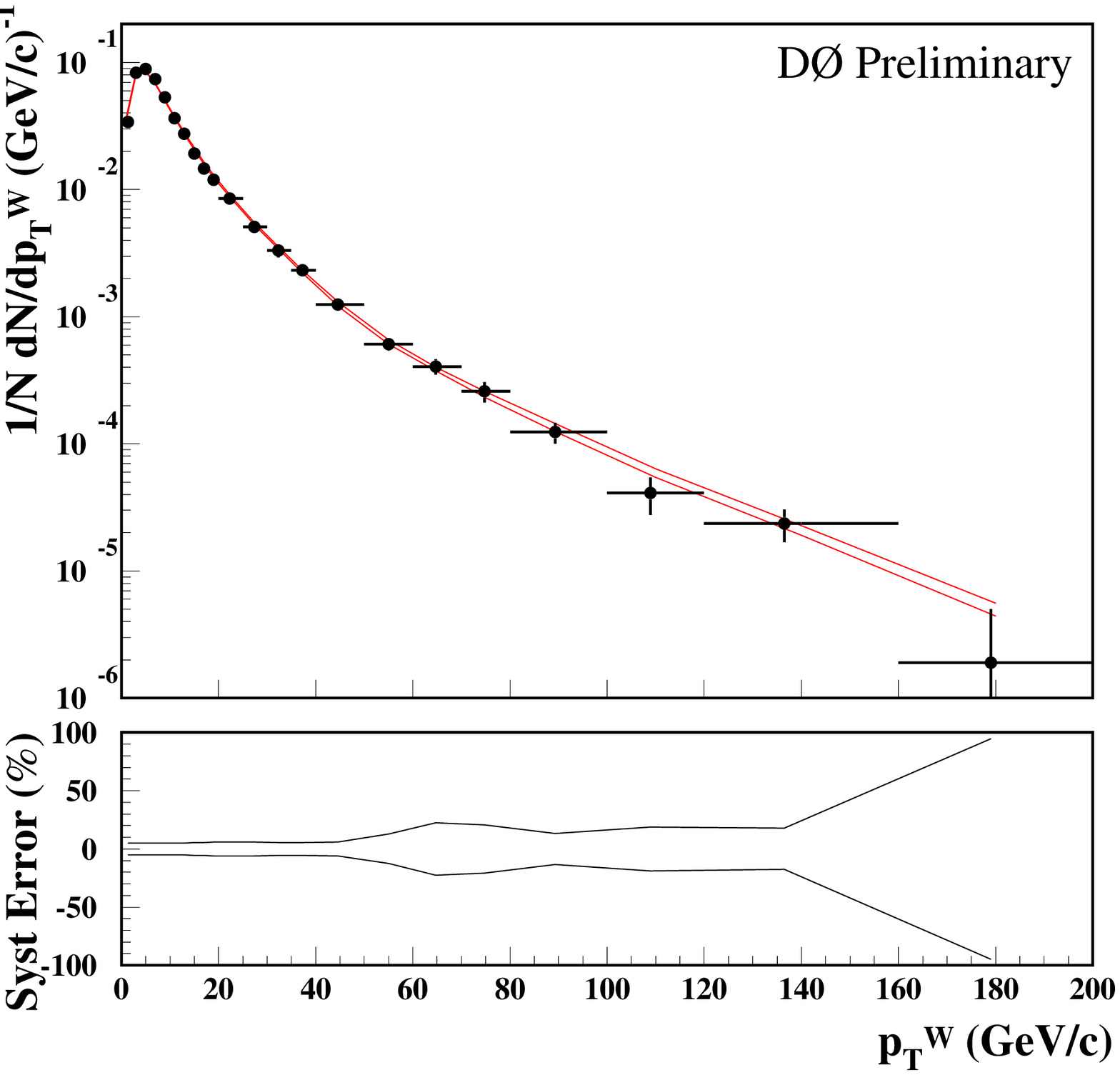,height=8.2cm,width=8.2cm}
&
\psfig{figure=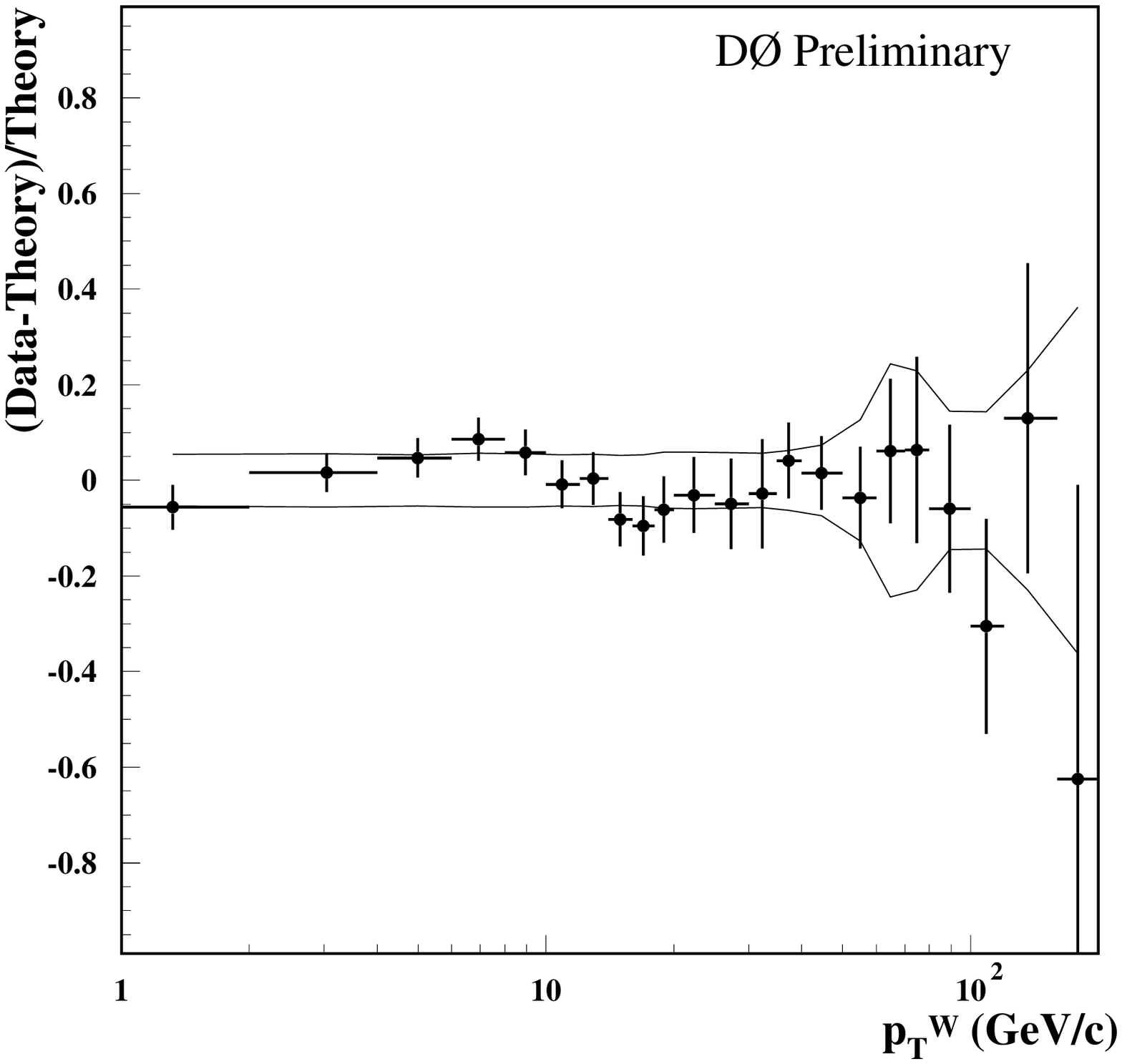,height=8.2cm,width=8.2cm}
\end{tabular}
\caption{ Left: D\O\ $p_T^{W}$ result (solid points) 
with statistical uncertainty.
The theoretical calculation by Arnold--Kauffman~\protect\cite{AKtheory}
has been smeared for detector
resolutions and is shown as two lines corresponding to the $\pm 1\sigma$
variations of the uncertainties in the detector modeling.
Data and theory are independently area normalized to unity.
The fractional systematic uncertainty on the data
is shown as a band in the lower portion of the plot.
Right: The ratio
(Data-Theory)/Theory shown as a function of $p_T^{W}$ with its statistical
uncertainty as error bars.
The total systematic uncertainties, shown as a band, were obtained by adding in quadrature
the contributions from data (background and efficiency)
and from the detector modeling.
}
\label{wpt}
\end{figure}

\begin{figure}[ht]
\begin{tabular}{l l}
\psfig{figure=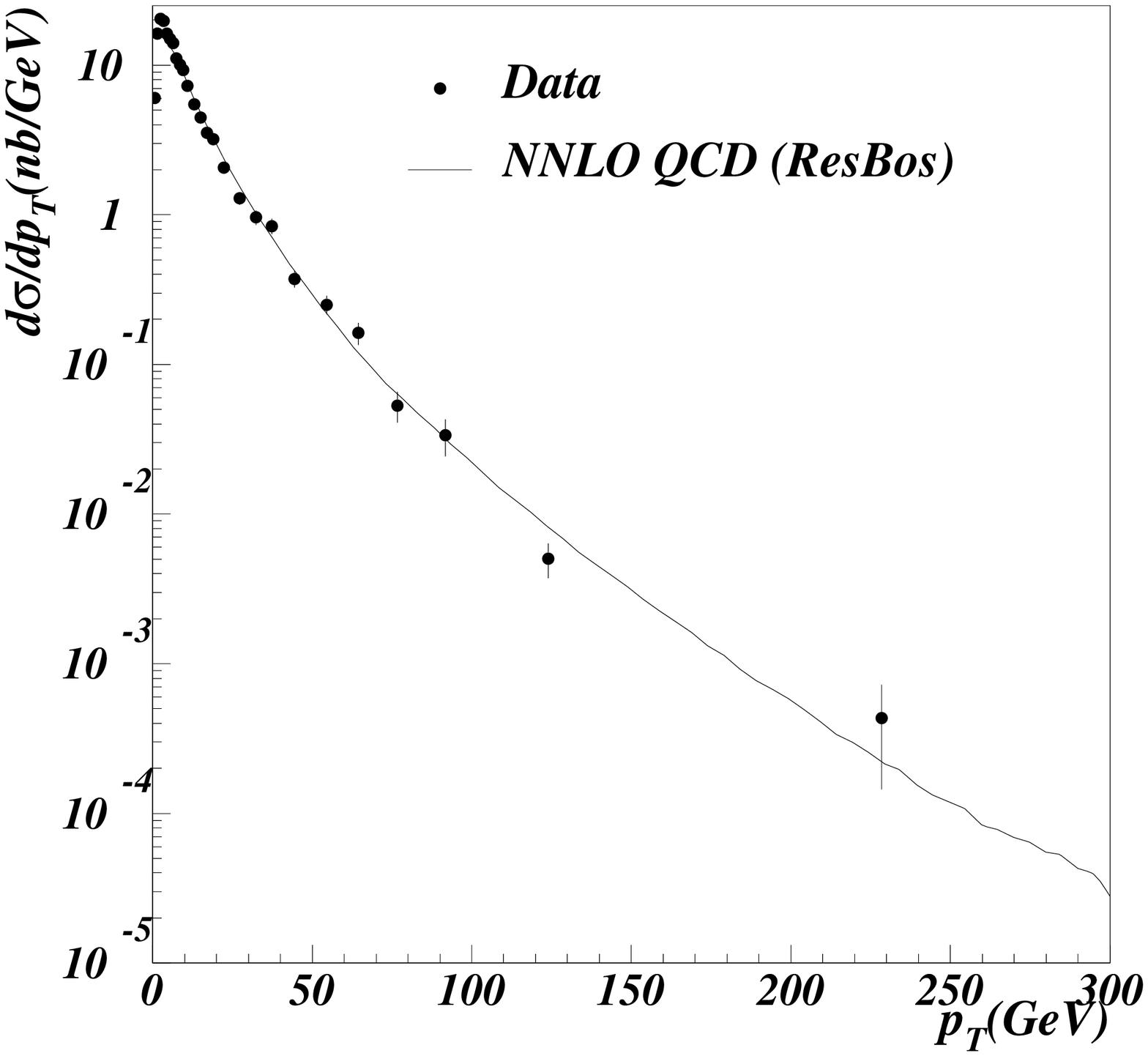,width=3.2in}
&
\psfig{figure=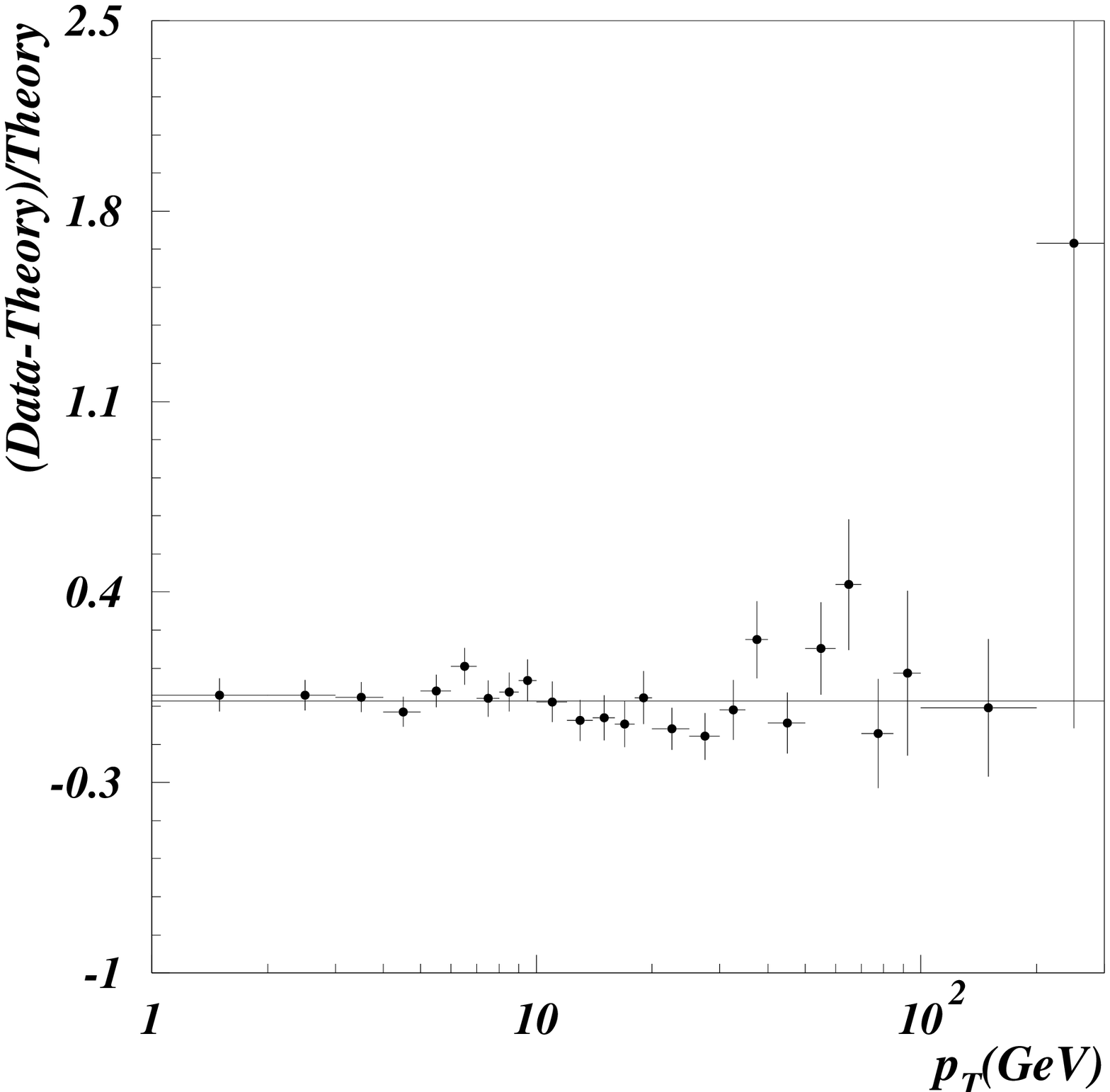,width=3.2in}
\end{tabular}
\caption{Left: \Dzero $Z$ differential cross section (circles) 
as a function of \pt compared to the calculation 
by Ladinsky--Yuan~\protect\cite{LY}.
Theory is normalized to the data.
Right: The ratio (Data-Theory)/Theory as a function of \pt of the $Z$.}
\label{zptres}
\end{figure}

\begin{figure}[ht]
\begin{tabular}{l l}
\psfig{figure=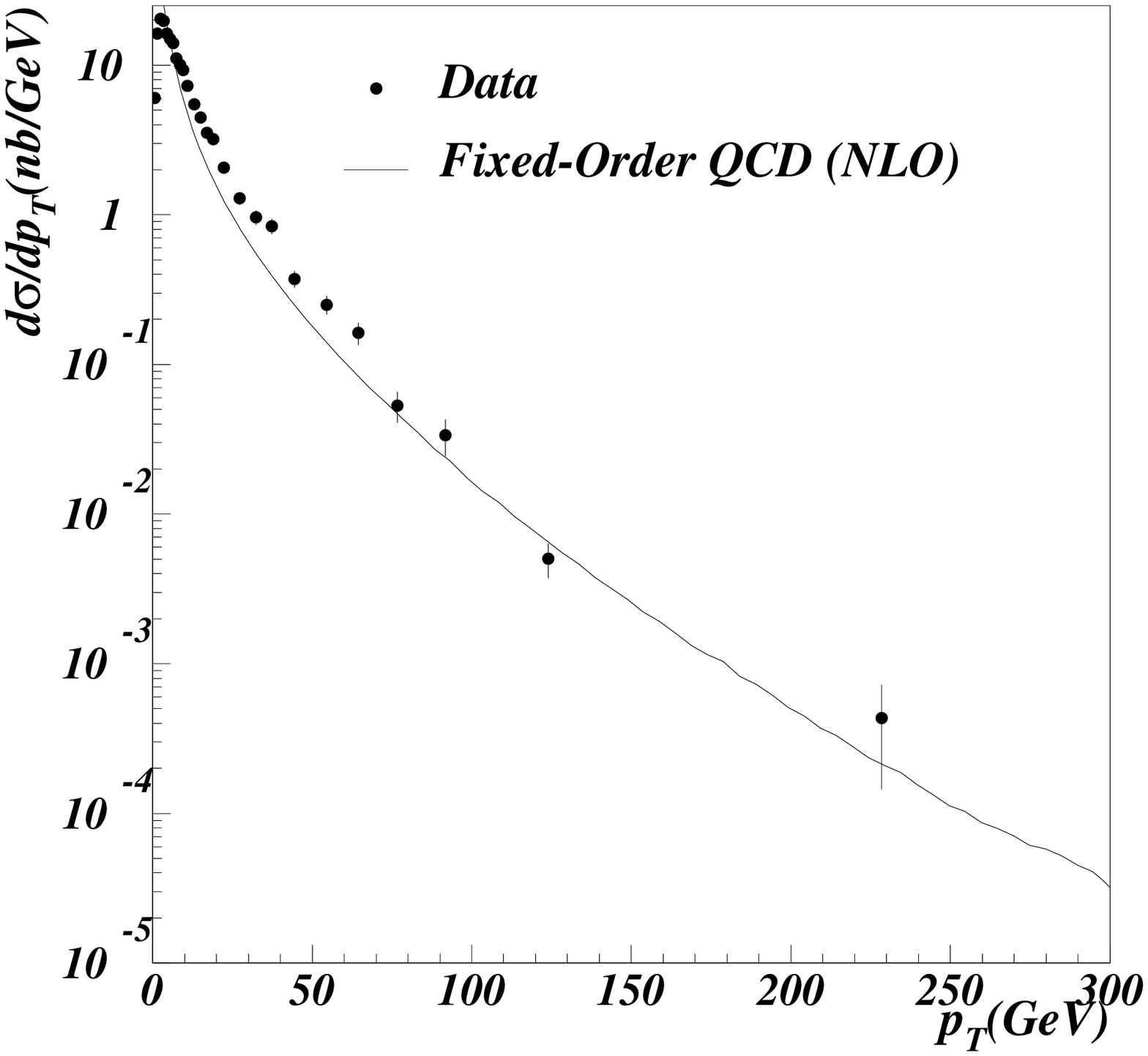,width=3.2in}
&
\psfig{figure=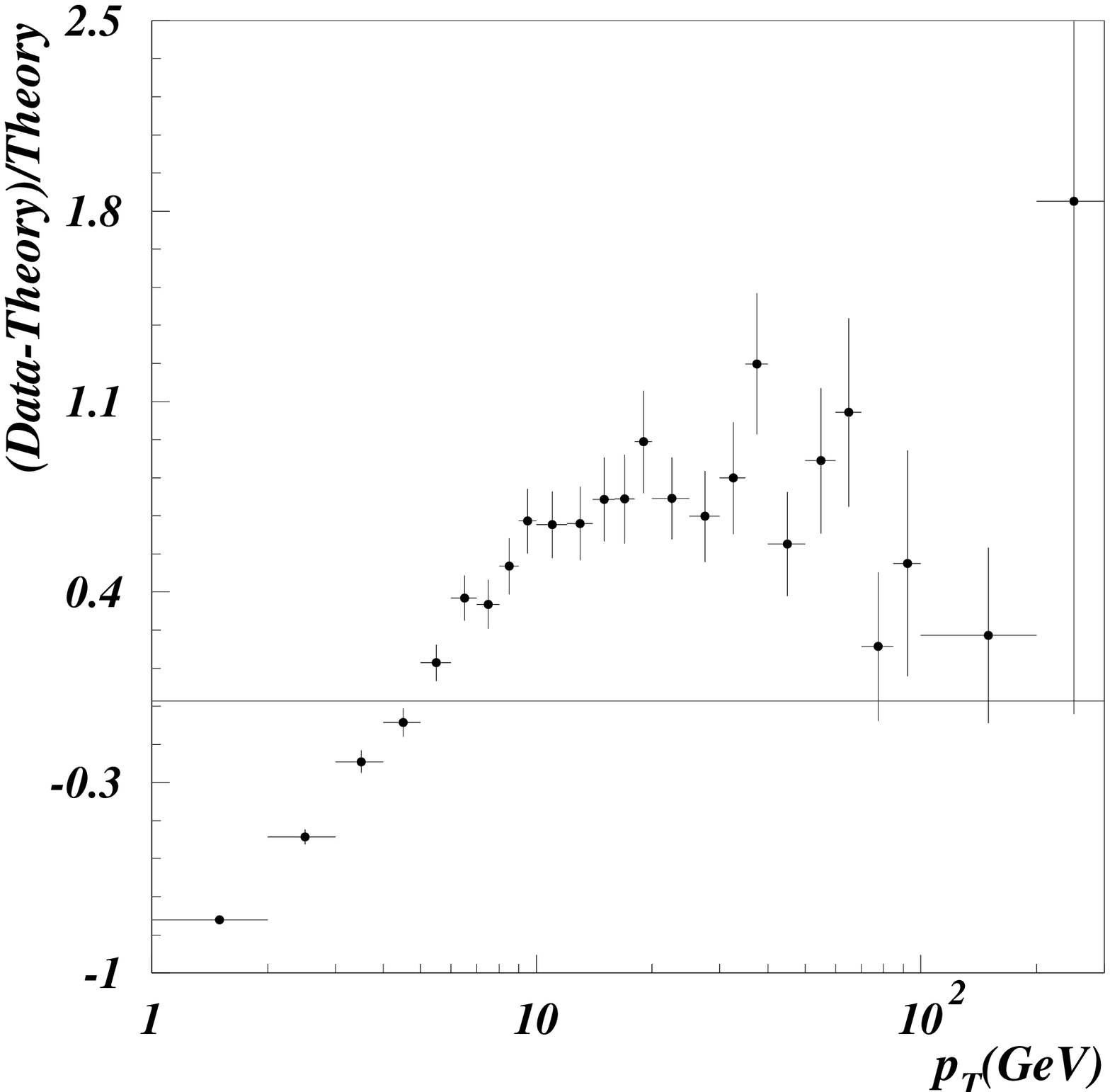,width=3.2in}
\end{tabular}
\caption{Left: \Dzero $Z~p_T$ smearing-corrected data (solid points) 
with total uncertainty
shown compared to the fixed-order (NLO) perturbative 
prediction~\protect\cite{ARtheory}.
The data is normalized to the \Dzero
measured inclusive \Z\ production cross section~\protect\cite{xsec};
the theory is normalized to its own prediction~\protect\cite{ARtheory}.
Right: Fractional difference between the \Z\ data and the fixed-order
calculation as a function of $p_T$.
}
\label{zptpert}
\end{figure}

The result for the $W$ \pt distribution, shown in Figure~\ref{wpt},
is compared to the theoretical calculation by Arnold and
Kauffman~\cite{AKtheory}, smeared by detector resolutions.
The $W$ data shows good
agreement with this combined QCD perturbative and resummation calculation
over the whole range of $p_T$.
In the case of the $Z$, we correct the measured cross section for the
effects of detector smearing.
Figure \ref{zptres} compares the final, smearing-corrected $Z$ \pt distribution 
to the calculation by Ladinsky and Yuan~\cite{LY}.
In addition, Figure~\ref{zptpert} compares the $p_T^Z$
measurement to the fixed-order perturbative theory~\cite{ARtheory}.
In comparing to the NLO calculation, we observed strong disagreement 
at low $p_T$, as expected due to the divergence of the calculation at $p_T=0$, 
and a significant enhancement of the cross section from soft gluon emission.

%
\clearpage

\section{Conclusions}
Using data taken with the \D0 detector during the 1994--1996 Tevatron collider
run, we have presented measurements of the 
$W$ and $Z$ transverse momentum distributions that 
agree well with the combined QCD perturbative and resummation calculations.

\section{Acknowledgments}

\input{acknowledgement_paragraph}


\end{document}

%% file: list_of_authors
%
B.~Abbott,$^{45}$                                                             
M.~Abolins,$^{42}$                                                            
V.~Abramov,$^{18}$                                                            
B.S.~Acharya,$^{11}$                                                          
I.~Adam,$^{44}$                                                               
D.L.~Adams,$^{54}$                                                            
M.~Adams,$^{28}$                                                              
S.~Ahn,$^{27}$                                                                
V.~Akimov,$^{16}$                                                             
G.A.~Alves,$^{2}$
% ...
\\                                                                            
\vskip 0.30cm                                                                 
\centerline{(D\O\ Collaboration)}                                             
\vskip 0.30cm                                                                 
\centerline{$^{1}$Universidad de Buenos Aires, Buenos Aires, Argentina}       
\centerline{$^{2}$LAFEX, Centro Brasileiro de Pesquisas F{\'\i}sicas,         
                  Rio de Janeiro, Brazil}                                     
\centerline{$^{3}$Universidade do Estado do Rio de Janeiro,                   
                  Rio de Janeiro, Brazil}                                     
\centerline{$^{4}$Institute of High Energy Physics, Beijing,                  
                  People's Republic of China}                                 
\centerline{$^{5}$Universidad de los Andes, Bogot\'{a}, Colombia}             
\centerline{$^{6}$Universidad San Francisco de Quito, Quito, Ecuador}         
\centerline{$^{7}$Institut des Sciences Nucl\'eaires, IN2P3-CNRS,             
                  Universite de Grenoble 1, Grenoble, France}                 
\centerline{$^{8}$DAPNIA/Service de Physique des Particules, CEA, Saclay,     
                  France}                                                     
\centerline{$^{9}$Panjab University, Chandigarh, India}                       
\centerline{$^{10}$Delhi University, Delhi, India}                            
\centerline{$^{11}$Tata Institute of Fundamental Research, Mumbai, India}     
\centerline{$^{12}$Kyungsung University, Pusan, Korea}                        
\centerline{$^{13}$Seoul National University, Seoul, Korea}                   
\centerline{$^{14}$CINVESTAV, Mexico City, Mexico}                            
\centerline{$^{15}$Institute of Nuclear Physics, Krak\'ow, Poland}            
\centerline{$^{16}$Institute for Theoretical and Experimental Physics,        
                   Moscow, Russia}                                            
\centerline{$^{17}$Moscow State University, Moscow, Russia}                   
\centerline{$^{18}$Institute for High Energy Physics, Protvino, Russia}       
\centerline{$^{19}$Lancaster University, Lancaster, United Kingdom}           
\centerline{$^{20}$University of Arizona, Tucson, Arizona 85721}              
\centerline{$^{21}$Lawrence Berkeley National Laboratory and University of    
                   California, Berkeley, California 94720}                    
\centerline{$^{22}$University of California, Davis, California 95616}         
\centerline{$^{23}$University of California, Irvine, California 92697}        
\centerline{$^{24}$University of California, Riverside, California 92521}     
\centerline{$^{25}$Florida State University, Tallahassee, Florida 32306}      
\centerline{$^{26}$University of Hawaii, Honolulu, Hawaii 96822}              
\centerline{$^{27}$Fermi National Accelerator Laboratory, Batavia,            
                   Illinois 60510}                                            
\centerline{$^{28}$University of Illinois at Chicago, Chicago,                
                   Illinois 60607}                                            
\centerline{$^{29}$Northern Illinois University, DeKalb, Illinois 60115}      
\centerline{$^{30}$Northwestern University, Evanston, Illinois 60208}         
\centerline{$^{31}$Indiana University, Bloomington, Indiana 47405}            
\centerline{$^{32}$University of Notre Dame, Notre Dame, Indiana 46556}       
\centerline{$^{33}$Purdue University, West Lafayette, Indiana 47907}          
\centerline{$^{34}$Iowa State University, Ames, Iowa 50011}                   
\centerline{$^{35}$University of Kansas, Lawrence, Kansas 66045}              
\centerline{$^{36}$Kansas State University, Manhattan, Kansas 66506}          
\centerline{$^{37}$Louisiana Tech University, Ruston, Louisiana 71272}        
\centerline{$^{38}$University of Maryland, College Park, Maryland 20742}      
\centerline{$^{39}$Boston University, Boston, Massachusetts 02215}            
\centerline{$^{40}$Northeastern University, Boston, Massachusetts 02115}      
\centerline{$^{41}$University of Michigan, Ann Arbor, Michigan 48109}         
\centerline{$^{42}$Michigan State University, East Lansing, Michigan 48824}   
\centerline{$^{43}$University of Nebraska, Lincoln, Nebraska 68588}           
\centerline{$^{44}$Columbia University, New York, New York 10027}             
\centerline{$^{45}$New York University, New York, New York 10003}             
\centerline{$^{46}$University of Rochester, Rochester, New York 14627}        
\centerline{$^{47}$State University of New York, Stony Brook,                 
                   New York 11794}                                            
\centerline{$^{48}$Brookhaven National Laboratory, Upton, New York 11973}     
\centerline{$^{49}$Langston University, Langston, Oklahoma 73050}             
\centerline{$^{50}$University of Oklahoma, Norman, Oklahoma 73019}            
\centerline{$^{51}$Brown University, Providence, Rhode Island 02912}          
\centerline{$^{52}$University of Texas, Arlington, Texas 76019}               
\centerline{$^{53}$Texas A\&M University, College Station, Texas 77843}       
\centerline{$^{54}$Rice University, Houston, Texas 77005}                     

%% file: acknowledgement_paragraph
%
We thank the Fermilab and collaborating institution staffs for
contributions to this work and acknowledge support from the 
Department of Energy and National Science Foundation (USA),  
Commissariat  \` a L'Energie Atomique (France), 
Ministry for Science and Technology and Ministry for Atomic 
   Energy (Russia),
CAPES and CNPq (Brazil),
Departments of Atomic Energy and Science and Education (India),
Colciencias (Colombia),
CONACyT (Mexico),
Ministry of Education and KOSEF (Korea),
and CONICET and UBACyT (Argentina).
%